\documentclass{PoS}
\usepackage{subfig}

\title{Nucleon and $N^* (1535)$ Distribution Amplitudes}

\ShortTitle{Nucleon and $N^* (1535)$ Distribution Amplitudes}

\author{V.~M.~Braun $^a$, S.~Collins $^a$, M.~G\"ockeler $^a$, C.~Hagen $^a$, R.~Horsley $^b$, Y.~Nakamura $^{a,c}$, D.~Pleiter $^d$, P.~E.~L.~Rakow $^e$, A.~Sch\"afer $^a$, \speaker{R.~W.~Schiel} $^{,a}$, G.~Schierholz $^f$, H.~St\"uben $^g$, J.~M.~Zanotti $^b$ \\
        \llap{$^a$} Institut f\"ur Theoretische Physik, Universit\"at Regensburg, 93040 Regensburg, Germany\\
        \llap{$^b$} School of Physics and Astronomy, University of Edinburgh, Edinburgh EH9 3JZ, UK \\
        \llap{$^c$} Center for Computational Sciences, University of Tsukuba, Tsukuba, Ibaraki 305-8577, Japan \\
        \llap{$^d$} John von Neumann Institute NIC / DESY Zeuthen, 15738 Zeuthen, Germany \\
        \llap{$^e$} Theoretical Physics Division, Department of Mathematical Sciences, University of Liverpool, Liverpool L69 3BX, UK \\
        \llap{$^f$} Deutsches Elektronen-Synchrotron DESY, 22603 Hamburg, Germany \\
        \llap{$^g$} Konrad-Zuse-Zentrum f\"ur Informationstechnik Berlin, 14195 Berlin, Germany \\
        E-mail: \email{rainer.schiel@physik.uni-regensburg.de}
        \begin{center} (The QCDSF Collaboration) \end{center}}

\abstract{The QCDSF collaboration has investigated the distribution amplitudes and wavefunction normalization constants of the nucleon and its parity partner, the $N^* (1535)$. We report on recent progress in the calculation of these quantities on configurations with two dynamical flavors of $\mathcal{O}(a)$-improved Wilson fermions. New data at pion masses of approximately 270 MeV helps in significantly reducing errors in the extrapolation to the physical point.}

\FullConference{The XXVIII International Symposium on Lattice Field Theory\\
		 June 14-19,2010\\
		 Villasimius, Sardinia Italy}

\begin{document}

\section{Introduction}

Distribution amplitudes (DAs) provide access to hadron wave functions and can be related to form factor data through perturbative QCD and light cone sum rules. This makes DAs interesting quantities to be investigated within the framework of lattice QCD. Distribution amplitudes for the nucleon and its parity-partner, the $N^* (1535)$, have been calculated by the QCDSF collaboration in \cite{Braun:2008ur, Braun:2009jy} on lattices with $N_f=2$ degenerate Clover fermions and pion masses of 400 to 1400 MeV. In this update, we present new data obtained for a $32^3 \times 64$-lattice with $\beta=5.29$, $\kappa=0.13632$ and 270 MeV pions. This new data is significantly closer to the physical point than before and will help us in reducing the uncertainties in the extrapolation to the physical point.

The nucleon distribution amplitudes are defined as follows \cite{Chernyak:1987nu}. In the infinite momentum frame, with transverse momentum components integrated out and only leading twist components considered, the nucleon wave function for the three-quark Fock state can be written as
\begin{equation}
| N, \uparrow \rangle = f_N \int \frac{ [dx] \varphi (x_i)}{2 \sqrt{24 x_1 x_2 x_3}} \{ | u^{\uparrow} (x_1) u^{\downarrow} (x_2) d^{\uparrow} (x_3) \rangle -  | u^{\uparrow} (x_1) d^{\downarrow} (x_2) u^{\uparrow} (x_3) \rangle \} 
\label{dadef}
\end{equation}
where $x_i$ are the longitudinal momentum fractions, the arrows indicate the nucleon and quark spins, $\int [dx] = \int_0^1 dx_1 dx_2 dx_3 \delta ( 1 - x_1 - x_2 - x_3 ) $, $f_N$ is the leading-twist normalization constant and $\varphi (x_i)$ is the nucleon distribution amplitude. We also calculate the normalization constants for the next-to-leading twist wave functions, $\lambda_1$ and $\lambda_2$; see Refs.~\cite{Braun:2008ur, Braun:2009jy} for the definition.

\section{Lattice procedure}

On the lattice, only moments of the distribution amplitude,
\[
\varphi^{lmn} = \int [dx] x_1^l x_2^m x_3^n \varphi (x_1, x_2, x_3) 
\]
can be accessed. And even this is limited to the first and second moments, since for higher moments, mixing with lower dimensional operators occurs, which would make the calculation increasingly difficult. Moreover, for higher moments, the matrix elements have to be evaluated at higher momenta which leads to noisier data. As can be seen in the following, even for the second moments we have barely enough statistics to get reasonably small error bars, and third moments would be much worse than that.

It is useful to expand the wave function in multiplicatively renormalizable terms \cite{Braun:2008ia}:
\begin{eqnarray*}
\varphi (x_i; \mu^2) & = & 120 x_1 x_2 x_3 \Big\{ 1 + c_{10} (x_1 - 2 x_2 +x_3) L^{\frac{8}{3\beta_0}} + c_{11} (x_1  - x_3) L^{\frac{20}{9\beta_0}} \\
& & + c_{20} \left[ 1 + 7 (x_2 - 2 x_1 x_3 - 2 x_2^2) \right] L^{\frac{14}{3\beta_0}} + c_{21}
\left( 1 - 4 x_2 \right) \left( x_1 - x_3 \right) L^{\frac{40}{9\beta_0}}
\\ & & \left. + c_{22}
\left[ 3 - 9 x_2 + 8 x_2^2 - 12 x_1 x_3 \right] L^{\frac{32}{9 \beta_0}}+\ldots
\right\}
\end{eqnarray*}
where $L\equiv \alpha_s(\mu)/\alpha_s(\mu_0)$ and $c_{ij}$ are the so-called ``shape parameters''. The $c_{ij}$ are given by linear combinations of the moments of the distribution amplitude $\varphi^{lmn}$ with $l + m + n \leq i$. Unlike the moments, which are constrained by the momentum conservation condition $\varphi^{lmn} = \varphi^{(l+1)mn} + \varphi^{l(m+1)n} + \varphi^{lm(n+1)}$, the $c_{ij}$ are independent nonperturbative parameters. In practice, they are obtained from the data on $\varphi^{lmn}$ using constrained fits.

The moments of the distribution amplitudes are calculated through matrix elements of the form
\[
\langle \mathcal{O} (x)_{\alpha \beta \gamma} \bar{\mathcal{N}} (y)_\tau \rangle
\]
where $\mathcal{N}$ is a smeared nucleon interpolator and $\mathcal{O}$ is a local three-quark operator with up to two derivatives. The operators $\mathcal{O}$ that are used to calculate the DAs have been classified according to irreducible representations of the lattice symmetry group in order to avoid mixing of operators \cite{Kaltenbrunner:2008pb} and they have been non-perturbatively renormalized \cite{G\"ockeler:2008we}.

\begin{figure}
\begin{center}
\subfloat[This effective mass plot shows a nice plateau for the nucleon (left, red line) and a reasonable plateau for the $N^* (1535)$ (right, green line).]{
\includegraphics[width=0.36\textwidth,angle=-90]{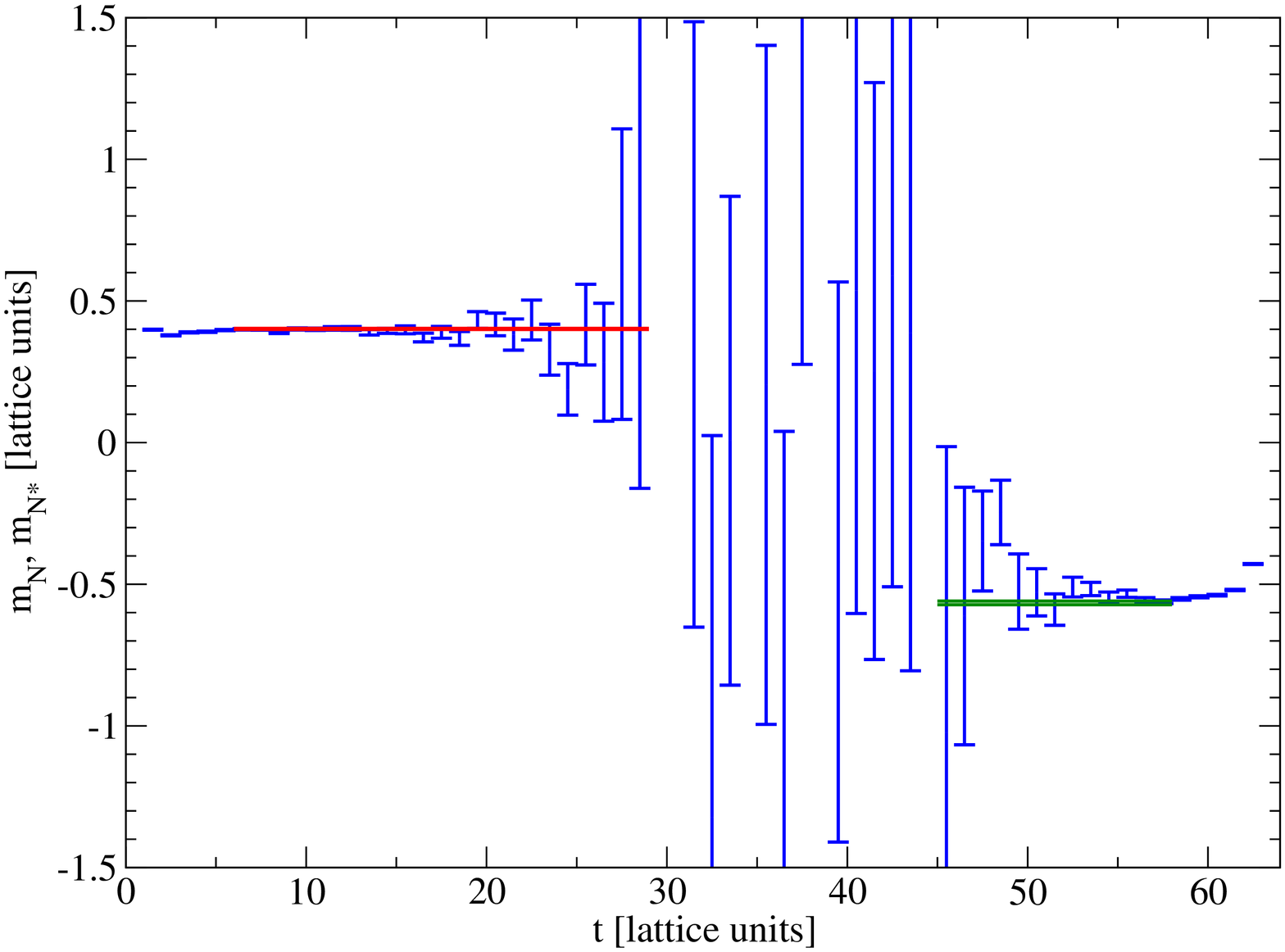} \label{effmass}}
\quad
\subfloat[The dispersion relation (dashed line: $E^2 = m^2 + p^2$) is fulfilled within the error bars. Shown here is data from six lattice ensembles with pion masses ranging from 270 MeV to 648 MeV.]{
\includegraphics[width=0.37\textwidth,angle=-90]{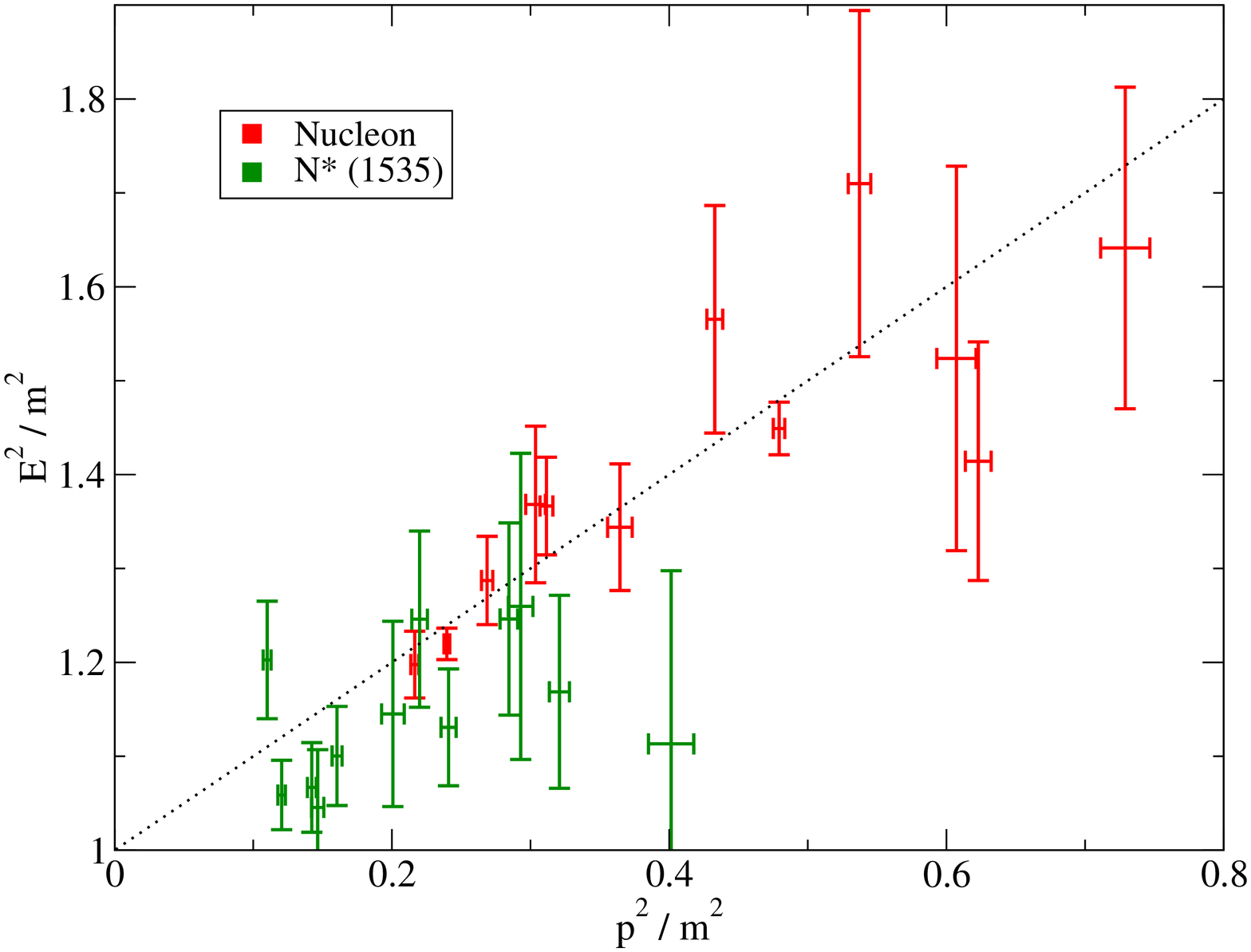} \label{disprel}}
\caption{Effective mass plot (a) and dispersion relation (b). Only statistical errors are shown.}
\end{center}
\end{figure}

Separation of the parity-plus (nucleon) and parity-minus ($N^* (1535)$) states has been achieved by using the generalized Lee-Leinweber parity projector (the non-generalized version of which was introduced in \cite{Lee:1998cx}), $\frac{1}{2} \Gamma \left( 1 + \frac{m}{E} \gamma_4 \right)$, where $\Gamma$ is a suitable product of $\gamma_i$ matrices and $m$ and $E$ are the mass and energy, respectively, of the baryon. Please note that this projector might have to be modified for certain operators and momenta, but it is perfectly suited for the operator-momentum combinations that are used here. In Fig.~\ref{effmass} one can see that a reasonable mass splitting is achieved between the nucleon and $N^*(1535)$ at momentum $p^2 = 0$. At higher momenta, the data becomes noisier, but it is still possible to find plateaus for both the nucleon and $N^*(1535)$. As a consistency check, the dispersion relation has been looked at and it is fulfilled within error bars (see Fig.~\ref{disprel}).

\section{Results}

\subsection{Normalization Constants}

\begin{figure}
\begin{center}
\subfloat{ 
\includegraphics[width=0.37\textwidth,angle=-90]{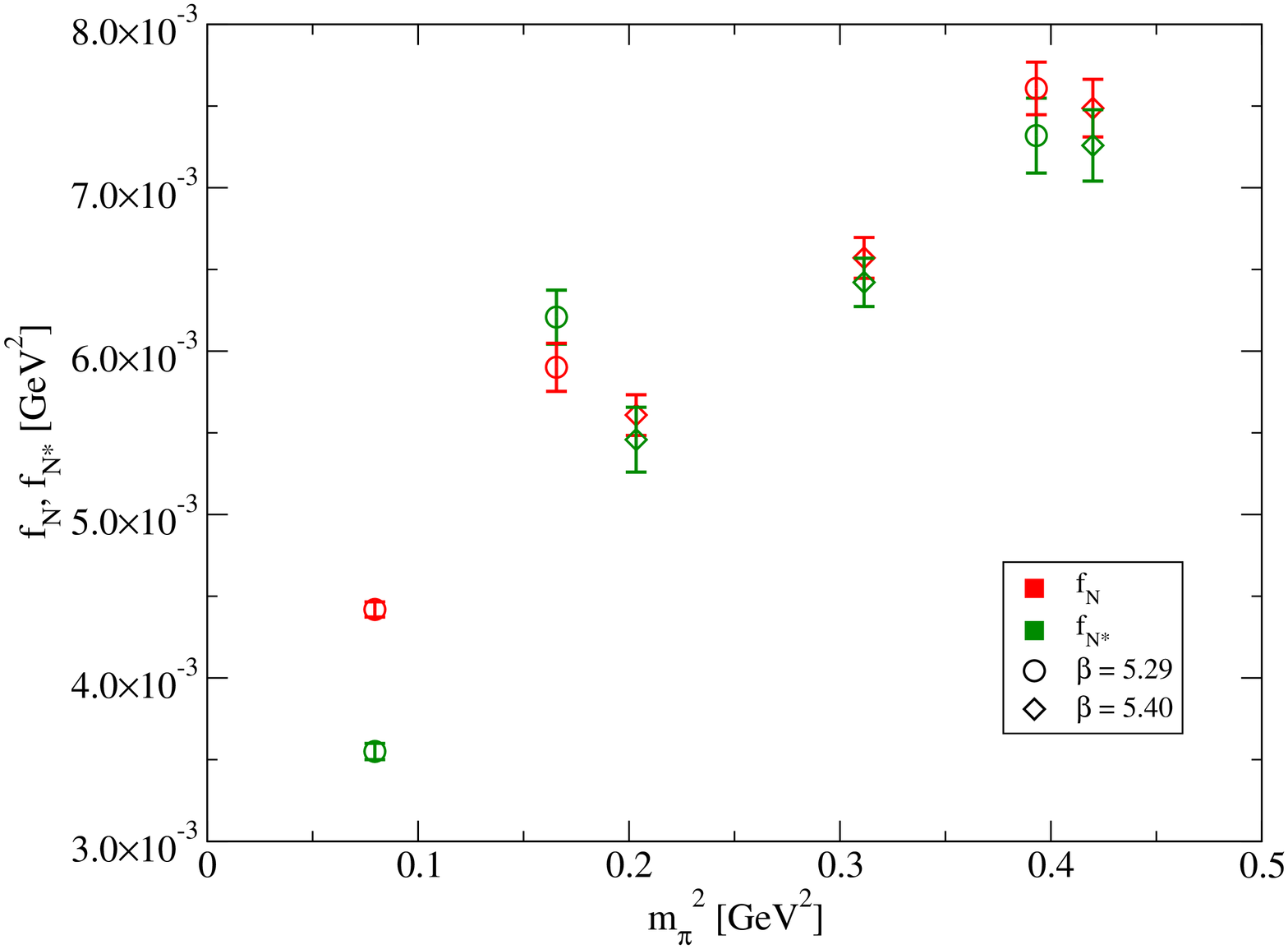}}
\subfloat{
\includegraphics[width=0.38\textwidth,angle=-90]{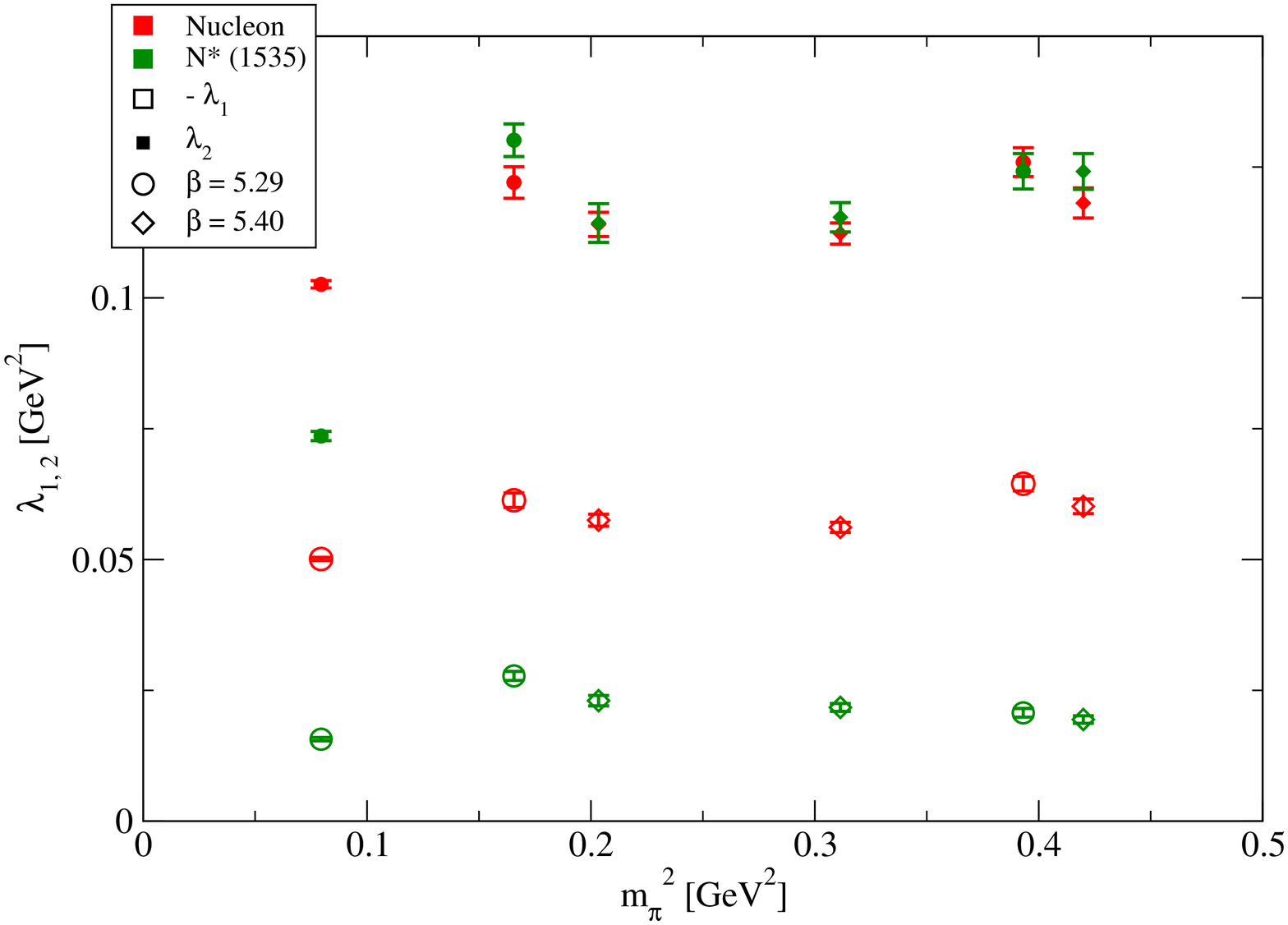}}
\end{center}
\caption{Dependence of $f_N$ and $f_{N^*}$ (left) and $\lambda_1$ and $\lambda_2$ (right) on the pion mass. Only statistical errors are shown.}
\label{resnorm}
\end{figure}

The results for the normalization constants $f_N$, $f_{N^*}$ and $\lambda_{1,2}$ are shown in Fig.~\ref{resnorm}. In comparison to the plots of the nucleon normalization constants shown in \cite{Braun:2008ur}, adding the new data point at 270 MeV pion mass makes a large difference, reducing the distance for the chiral extrapolation by more than a factor of two. Thus we can avoid using data with pion masses $> 1$ GeV which strongly influenced our previous results \cite{Braun:2008ur}. 

It is interesting to note that on the lattice with 270 MeV pion mass, it is the first time to see a significant difference between the leading-twist normalization constants of the nucleon and $N^*(1535)$. While $f_N$ and $f_{N^*}$ agree -- within error bars -- on the lattices with heavier pions, this is no longer true for the new data point. However, we do not yet have an explanation why this is the case or whether it should have been expected.

\subsection{Shape parameters}

\begin{figure}
\begin{center}
\includegraphics[width=0.50\textwidth,angle=-90]{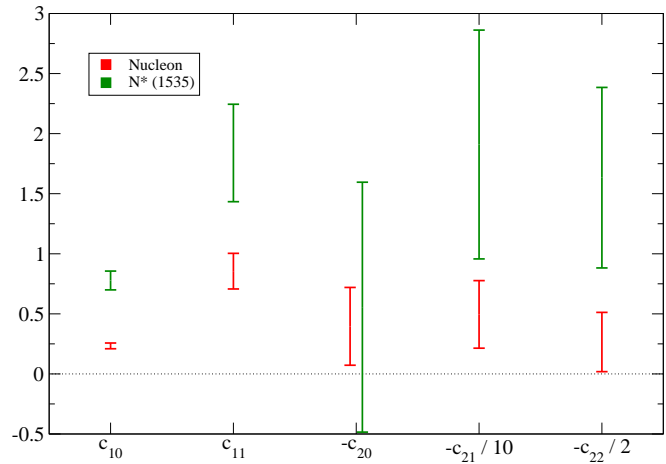}
\caption{Shape parameters for the nucleon and $N^*(1535)$ on the $32^3 \times 64$ lattice at 270 MeV pion mass. Only statistical errors are shown. While the error bars for the first moments are fine, more statistics will improve the errors bars for the second moments.}
\label{shapeparams}
\end{center}
\end{figure}

The shape parameters are calculated from the moments of the DAs as described above. Since the shape parameters describe the deviation from a symmetric momentum distribution, their relativ error is larger than the error of the moments of the DAs. Also, as can clearly be seen in Fig.~\ref{shapeparams}, the second order shape parameters $c_{2j}$ are noisier than the first order ones, $c_{1j}$. This can be mostly attributed to the fact that the calculation of the $c_{2j}$ involves matrix elements at $p^2 = 2$, while one needs only the cleaner $p^2=1$ matrix elements to calculate the $c_{1j}$. Also, the nucleon error bars are smaller than the $N^*(1535)$'s, which is not surprising given that the $N^*(1535)$ has a larger mass and thus a shorter plateau than the nucleon.

\begin{figure}
\begin{center}
  \includegraphics[width=0.75\textwidth]{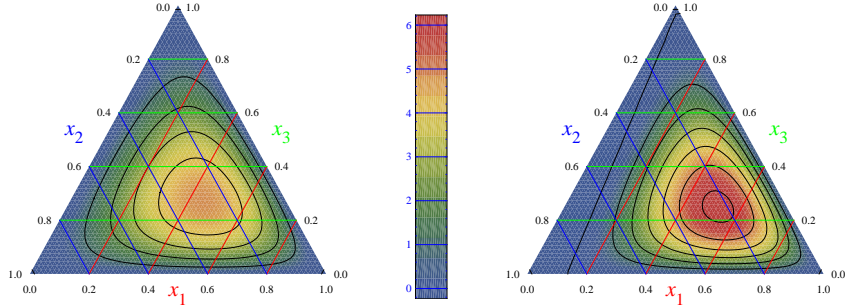}
  \caption{Barycentric plot of the nucleon (left) and $N^* (1535)$ (right) distribution amplitudes, calculated for the central values of the shape parameters. Due to the large error bars on the second moments, only the first moments have been used for this plot.}
\label{barycplot}
\end{center}
\end{figure}

The first order shape parameters show a significant difference between the nucleon and its parity partner. This difference is also visualized in the barycentric plot, Fig.~\ref{barycplot}: The $N^*(1535)$ wavefunction is much more strongly peaked towards the spin-carrying quark (quark label ``1'' in Eq.~\ref{dadef}) than the nucleon.

The second order shape parameters still have very large error bars and are thus not included in the barycentric plot. However, we hope that increased statistics will reduce these error bars to a reasonable size. While $c_{20}$ is compatible with 0 both for the nucleon and its parity partner, $c_{21}$ and $c_{22}$ tend to be negative in both cases.

\section{Conclusions and Outlook}

Nucleon and $N^*(1535)$ distribution amplitudes and the wavefunction normalization constants $f_N$, $f_{N^*}$ and $\lambda_{1,2}$ have been calculated at pion masses as low as 270 MeV. Unlike for higher pion mass lattices, a difference between the nucleon and $N^*(1535)$ wavefunction normalization constants can now be seen. The distribution amplitudes for the nucleon and its parity partner are also significantly different, with the $N^*(1535)$'s wavefunction being more peaked towards the spin-carrying quark than the nucleon's wavefunction.

To improve the accuracy of the results on the nucleon and $N^*(1535)$ DAs, more statistics are being collected for the $32^3 \times 64$ lattice with 270 MeV pions. Also, calculations at the same pion mass but a larger lattice volume will be performed to get a handle on finite volume effects. Finally, the data will have to be extrapolated to the physical point.

\begin{acknowledgments}

This work has been supported in part by the Deutsche Forschungsgemeinschaft (Sonderforschungsbereich / Transregio 55) and the Research Executive Agency (REA) of the European Union under Grant Agreement number PITN-GA-2009-238353 (ITN STRONGnet). The computations were performed on Regensburg's Athene HPC cluster using the Chroma software system \cite{Edwards:2004sx}.

\end{acknowledgments}

\end{document}